\documentclass[reprint,aps,prl]{revtex4-1}

\usepackage{graphicx}
\usepackage{epsfig}
\usepackage{epstopdf}
\usepackage{subfigure,color}
\usepackage{dcolumn}
\usepackage{bm}
\usepackage{amsmath,amssymb}

\def\.{\cdot}

\def\##1{{\bf #1\mit}}
\def\_#1{{\bf #1\mit}}
\def\-#1{{\bf #1\mit}}
\def\=#1{\overline{\overline #1}}

\begin{document}
	\newcommand{\red}[1]{\textcolor{red}{#1}}
	\title{Nonreciprocity in Bianisotropic Systems with Uniform Time Modulation}
	
	\author{X.~Wang$^{1}$}\email{These authors contributed equally.}
	\author{G.~Ptitcyn$^{1}$}\email{These authors contributed equally.}
	\author{V.~S.~Asadchy$^2$}\email{These authors contributed equally.}
	\author{A.~D\'{i}az-Rubio$^1$}
	\author{M.~S.~Mirmoosa$^3$}
	\author{Shanhui~Fan$^2$}
	\author{S.~A.~Tretyakov$^1$}
	\affiliation{$^1$Department of Electronics and Nanoengineering, Aalto University, P.O.~Box 15500, FI-00076 Aalto, Finland\\
		$^2$Ginzton Laboratory and Department of Electrical Engineering, Stanford University, Stanford, California 94305,  USA\\
		$^3$Laboratory of Wave Engineering, Swiss Federal Institute of Technology in Lausanne (EPFL), CH-1015 Lausanne, Switzerland} 
	
	\begin{abstract}
		Physical systems with material properties   modulated in time  provide versatile routes for  designing magnetless   nonreciprocal devices. Traditionally, nonreciprocity in such systems is achieved exploiting both temporal and spatial modulations, which inevitably requires a series of time-modulated elements distributed in space.   In this paper, we introduce a concept of bianisotropic time-modulated systems capable of nonreciprocal wave propagation at the fundamental frequency and based on  uniform, solely temporal  material  modulations. In the absence of temporal modulations, the considered bianisotropic systems are reciprocal.
		We theoretically explain the nonreciprocal effect by analyzing wave propagation in an unbounded bianisotropic time-modulated medium. The effect stems 
			from temporal modulation of spatial dispersion effects which  to   date  were not taken   into account in previous studies based on the local-permittivity description.
		We propose  a    circuit design of a bianisotropic metasurface that can provide phase-insensitive isolation and unidirectional amplification.
	\end{abstract}
	\maketitle

	Reciprocity is a fundamental principle of a physical system,  requiring that the transmission between two ports does not change if the source and receiver are interchanged. 
	Breaking reciprocity is necessary for unidirectional wave propagation, such as wave isolation and circulation~\cite{caloz_electromagnetic_2018,asadchy_tutorial_2020}.
	The conventional way for attaining nonreciprocity is to exert magnetic bias on magneto-optical materials~\cite{WangFan2005,Bi2011}, which however has rather weak effect at high frequencies. Moreover, the devices based on magneto-optical materials are bulky and incompatible to systems where parasitic effects of external magnetic fields should be avoided. 
	An alternative approach is to use nonlinear materials~\cite{lepri2011,soljavcic2003,shadrivov2011,Lfan2012}, but it only works for certain strengths of the incident signal and the functionalities are limited by the dynamic reciprocity constraint~\cite{shi_limitations_2015}.
	

	Dynamic modulation of the material properties brings an additional degree of freedom for obtaining unprecedented wave effects in acoustics~\cite{fleury2014,trainiti2019}, optics~\cite{sounas2014,phare_Lipson2015},   and microwave engineering~\cite{estep2014magnetic,mirmoosa2019time,ptitcyn2019time}. 
	It was  noticed quite early that  an electronic device  whose properties are modulated in space and in time    can exhibit nonreciprocal response~\cite{Cullen1958,Kamal,wentz1966}.
	In the last decade, due to advances in electronics and photonics, research interest to nonreciprocal wave propagation based on space-time modulated systems has rapidly revived and yielded various  designs of nonreciprocal devices: isolators \cite{bhandare2005, yuFAN2009,fang2012photonic, LiraLipson2012,correas2015}, circulators \cite{shi2017optical}, phase-shifters \cite{yuFan2009opt,wang2020theory}, and one-way amplifiers~\cite{abdo2013directional,song2019direction,galiffi2019broadband}.
	To date, all the known approaches for obtaining nonreciprocal wave propagation in time-modulated systems can be boiled down to the following three fundamental classes~\cite{sounas2017non,williamson2020integrated}: Travelling-wave modulators (indirect photonic transitions)~\cite{Cullen1958,dong_inducing_2008,yu_complete_2009,song2019direction,wang2020theory,qin_nonreciprocal_2014,hadad_breaking_2016,taravati2020space,taravati2017nonreciprocal,chamanara2017optical,shaltout2019spatiotemporal,shaltout2015time,cardin2020surface,zhang2019breaking,li2020time,taravati2019generalized,ramaccia2018nonreciprocity,taravati2017self,taravati2019dynamic},  tandem phase modulators (and related approach based on direct photonic transitions)~\cite{doerr_optical_2011,fang2012photonic,galland_broadband_2013,baldwin1961nonreciprocal,hamasaki1964theory,brenner1967unilateral}, and nonreciprocal frequency converters~\cite{koutserimpas2018nonreciprocal,wu_serrodyne_2019,helming1964non}. The first   approach implies modulation of material properties in both  space and in time, while the second  requires  two temporally  modulated components separated in space. In both cases, it is necessary  to   use   a series of time-modulated elements which have to be precisely synchronized with each other, which greatly increases the complexity of the biasing networks. The third approach requires either   asymmetric modulation function profile of the real part of permittivity~\cite{wu_serrodyne_2019,williamson2020integrated} or modulating both its real and imaginary parts~\cite{koutserimpas2018nonreciprocal}. However, in both cases the system  exhibits reciprocal transmission for the fundamental frequency   since   waves incident from the opposite directions ``sense'' effectively the same structure (nonreciprocity manifests itself only in nonreciprocal frequency  conversion). Consequently, designing isolators using this   frequency-converter approach   requires cascading a pair of two converters, which   results in additional device complexity~\cite{koutserimpas2018nonreciprocal,wu_serrodyne_2019}.  
	
	In this paper, we introduce a  concept of linear bianisotropic  time-modulated systems capable of nonreciprocal wave propagation at  the fundamental frequency and implying solely temporal and uniform modulation of material properties.  
	This route for nonreciprocal time-modulated systems,   originated from  bianisotropy (weak spatial dispersion),  strikingly differs from the previously known three approaches based on the local-permittivity material description.
It should be mentioned that in nonlinear systems,  nonreciprocal response under  uniform temporal modulation  is possible by creating an external angular-momentum bias~\cite{duggan_optically_2019}. 
	In addition to the fundamental theoretical importance, our approach additionally provides certain advantages 
	for practical realization (it is sufficient to ensure temporal modulation of a single component in the nonreciprocal system).
	We explain and demonstrate the physics behind the new effect by analyzing 
	wave propagation in an unbounded bianisotropic time-modulated medium (such a medium is reciprocal in the absence of temporal modulations).  Next, we extend the study to two-dimensional bianisotropic metasurfaces  (single-layer metamaterial composites).  
	We design a deeply sub-wavelength   metasurface which exhibits  strong unidirectional transmission or unidirectional  amplification. The metasurface incorporates a single temporally modulated capacitive layer backed by a usual dielectric layer.  
	We show that the metasurface  obeys the generalized time-reversal symmetry, but exhibits strong  unidirectional amplification/attenuation.   Finally, we propose an equivalent circuit for the bianisotropic metasurface capable of phase-insensitive isolation.

	First, we analyze   wave propagation in unbounded materials whose effective material parameters are modulated in time according to the same symmetric profile and with the same phase at each point in space (uniform or so-called global modulation). It will be shown that wave propagation in arbitrary anisotropic materials with global time modulation is always reciprocal. On the other hand, it will be shown that under the same conditions,  reciprocity can be broken in bianisotropic materials.
	
	The constitutive relations of a   bulk bianisotropic material (reciprocal in the absence of temporal modulations) with antisymmetric magnetoelectric tensor (describing so-called omega magnetoelectric coupling) can be written in the form of~\cite[Eq.~8.4]{serdyukov_electromagnetics_2001}
	\begin{equation}
		\_D = \=\varepsilon \cdot \_E + \Omega \=J \cdot \_H, \quad 
		\_B = \=\mu \cdot \_H + \Omega \=J \cdot \_E, \label{bian11}
	\end{equation}
	where $\=\varepsilon$ and $\=\mu$ are the anisotropic permittivity and permeability tensors, $\Omega$ is the amplitude of  the bianisotropic omega coupling, and $\=J=\hat{\_z} \times \=I$ is the transverse vector-product dyadic. 
	Here, for simplicity, we use the adiabatic model for temporal modulations, assuming that the operational frequency~$\omega$  is very low compared to the lowest resonance frequency of the material. 
	In this case,   the uniformly modulated material tensors  can be written as~\cite[\textsection~1]{suppl}, $\=\varepsilon (\omega,t,\_r) =  \=\varepsilon_{\rm st}(\omega,\_r)+ \=M_\varepsilon (\_r) \cos (\omega_{\rm m} t + \phi)$, $\=\mu (\omega,t,\_r) =  \=\mu_{\rm st}(\omega,\_r)+ \=M_\mu (\_r) \cos (\omega_{\rm m} t + \phi)$, and $\Omega (\omega,t,\_r) =   M_\Omega (\_r) \cos (\omega_{ \rm m} t + \phi)$,
	where $\=\varepsilon_{\rm st}$ and $\=\mu_{\rm st}$ denote static (in the absence of time modulation) permittivity and permeability, $\=M_\varepsilon$, $\=M_\mu$, and $M_\Omega$ are  the modulation
	strength functions, $\omega_{\rm m}$ is the modulation frequency, and $\phi$ is an arbitrary global phase. 
	Note that this model   can be used for arbitrary  modulation frequency~$\omega_{\rm m}$ (see~\cite[\textsection~1]{suppl} for details). In the general non-adiabatic case,  the following derivations  could still be performed, writing  the material parameters   using  integrals over past time.
	\nocite{ohler_electromagnetic_1999,mirmoosa_dipole_2020, martin2015artificial,radi_tailoring_2014,yazdi2015bianisotropic,alaee_all-dielectric_2015,slobozhanyuk_three-dimensional_2017,9063633,li2019transfer,tretyakov_analytical_2003,pozar_microwave_2012}
	
	Due to the periodical modulation, the electric and magnetic fields  are written in terms of the Fourier components at frequencies $\omega_n=\omega_0 +n \omega_{\rm m}$, i.e. $\_E_n$ and $\_H_n$. The external sources are characterized by the electric current harmonics $\_J_{ e,n}$, and time-harmonic oscillations in the form ${\rm e}^{+j \omega t}$ are assumed.
	Analogously to derivations in~\cite{shi_multi-frequency_2016},  Eqs.~(\ref{bian11})  can be substituted into  Maxwell equations and the wave equation can be written in the matrix form \cite[\textsection~2]{suppl},
	\textcolor{black}{
	\begin{equation}
	\begin{array}{ll}\displaystyle
	\hspace*{-0.3cm}  -j \left( \left\{ j  \Big[\omega \Big]^{-1}\cdot\Big[D \Big]  +    \Big[A_\Omega\Big]\cdot\Big[J \Big] \right\} \cdot \Big[A_\mu\Big]^{-1}  \right.
	\vspace*{.2cm}\\\displaystyle
	\left. \cdot \left\{ j  \Big[\omega\Big]^{-1}\cdot\Big[D \Big]  - \Big[A_\Omega\Big]  \cdot\Big[J \Big]    \right\}  + \Big[A_\varepsilon\Big]   \right) \cdot \Big[\_E\Big] =\Big[\_J_{e}'\Big].
	\end{array}
	\label{bian4}
	\end{equation}}
	Here,  $\Big[\_E\Big]$ and $\Big[\_J_{ e}'\Big]$ denote the column vectors with components $\_E_n$ and $\_J_{e,n}'=\_J_{e,n}/\omega_n$, and $\Big[\omega \Big]$ is the diagonal matrix with frequencies $\omega_n$ at the diagonal. Block matrix $\Big[D \Big]$ denotes vector operation $\nabla\times$ and block matrix $\Big[J \Big]$ is composed of antisymmetric matrices $\=J$ on the diagonal (see the definitions of the matrices in \cite[\textsection~2]{suppl}). Block matrices $\Big[A_\varepsilon\Big]$, $\Big[A_\mu\Big]$, and $\Big[A_\Omega\Big]$ are described in~\cite[\textsection~2]{suppl} and include dependence on the modulation strength functions $\=M_\varepsilon$, $\=M_\mu$, and $M_\Omega$, respectively. They can be made symmetric by selecting global phase $\phi=0$ (the initial phase can be chosen arbitrarily by time translation $t \rightarrow t +\Delta t$). Equation~(\ref{bian4}) can be simplified to $\Big[\_E\Big] = \Big[G\Big] \cdot \Big[\_J_{e}'\Big]$ with $ \Big[G\Big]$ being  Green’s function of the time-varying unbounded material written in the block matrix form.
	\begin{figure*}
		\centering
		\includegraphics[width=0.95\linewidth]{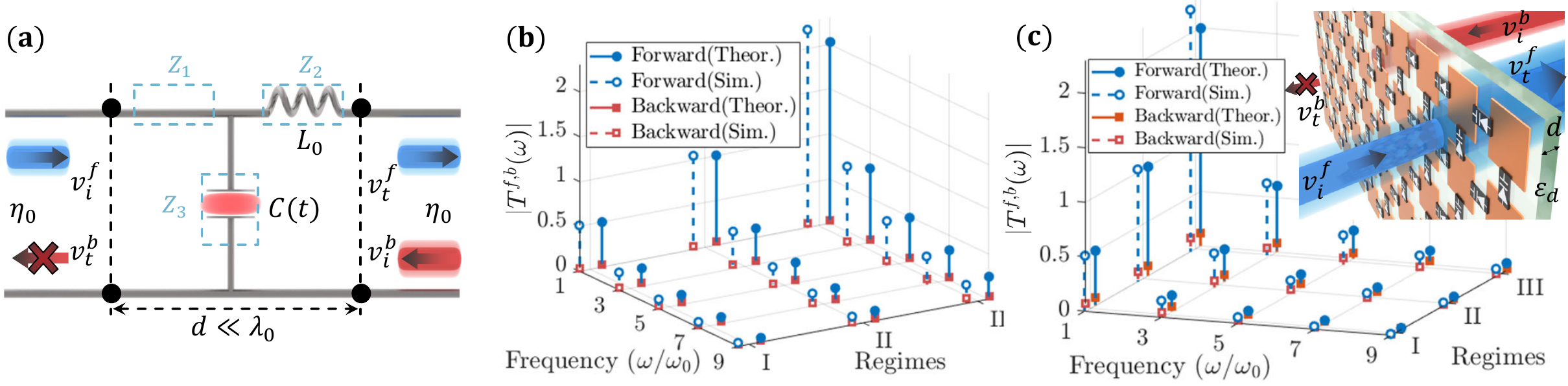}
		\caption{ (a) A T-circuit which  describes  propagation of plane waves through a bianisotropic time-modulated metasurface. 
			(b) Spectral response of the time-varying $LC$- circuit for forward and backward incident waves.
			In all three regimes, $L_0=6.87$~nH, $\phi=0$, and $A=1$. The capacitances $C_0$  for the three regimes are: $C_0=2.2$~pF for regime~I, $C_0=1.6$~pF for regime~II, and $C_0=1.55$~pF for regime~III. 
			The operating frequency is $f_0=10$~GHz. 
		 (c) Spectral response of the metasurface. Here, $C(t)=6.8[1-0.9\sin(\omega_{m} t+0.084\pi)]$~pF for regime I, $C(t)=5.24[1-0.9\sin(\omega_{m} t+1.131\pi)]$~pF for regime II,  $C(t)=4.12[1-0.9\sin(\omega_{ m}t+1.195\pi)]$~pF  for regime III, $f_0=10$~GHz, $\epsilon_{ d}=65$, and $d=\lambda_0/30$. } \label{fig_circuit}
	\end{figure*}
As shown in \cite[\textsection~2]{suppl},  Green's function matrix of 
	any anisotropic  material (i.e., when $A_\Omega=0$) is symmetric~\cite{fang2012photonic}, which implies that wave propagation in such
	material  is reciprocal and subject to the Lorentz reciprocity~\cite[Eq.~119]{asadchy_tutorial_2020}. On the contrary, an unbounded bianisotropic omega material with nonzero $A_\Omega$ breaks reciprocity since in this case matrix $\Big[G\Big]$ is always not symmetric. It should be noted that nonreciprocal transmission in bianisotropic material occurs even when permittivity and permeability are time-invariant, i.e. $  \= \varepsilon (\omega,t,\_r) =  \=\varepsilon_{\rm st}(\omega,\_r)$ and $  \= \mu (\omega,t,\_r) =  \=\mu_{\rm st}(\omega,\_r)$. It is important to mention that nonreciprocity   requires antisymmetric magnetoelectric coupling and cannot be achieved in isotropic chiral materials with globally  modulated properties. This can be easily verified  replacing $\Omega \=J$ in the former of (\ref{bian11}) by $-\kappa \=I$ and  the latter  by $+\kappa \=I$~\cite[Eq.~8.4]{serdyukov_electromagnetics_2001}.

The above derivations demonstrate that  a bulk  material with temporally modulated bianisotropic response supports nonreciprocal wave propagation. Since in most practical situations implementation of bulk bianisotropic materials can be complicated in terms of fabrication, next we consider the same effect in a two-dimensional single-layer array of bianisotropic elements (a metasurface).
Analogously to bulk materials which can be modeled by volume-averaged material parameters, metasurfaces are conventionally characterized by surface-averaged material parameters, i.e. polarizabilities, susceptibilities, or surface impedances~\cite[\textsection~2.4]{yang_rahmat-samii_2019}. Thus, the above conclusions for time-varying bulk materials also apply to the time-modulated metasurfaces.
In what follows, we choose the surface impedance model,  which represents a 
metasurface as an equivalent  circuit of specific configuration. Propagating plane waves with electric~$E$ and magnetic~$H$ fields are modeled by signals  with voltages~$v$ and currents~$i$  propagating in an equivalent transmission line~\cite[\textsection~3]{suppl}. 

	Any reciprocal bianisotropic metasurface can be described  by an equivalent T- or $\Pi$-circuit. 
	We model a  metasurface with a T-circuit formed by three lumped impedances in frequency domain, $Z_1$ and $Z_2$ connected in series and $Z_3$ connected in parallel, as shown in Fig.~\ref{fig_circuit}(a). The total thickness of the metasurface~$d$ can be deeply sub-wavelength. 
	In such representation, the series impedances characterize effective magnetic polarization in the metasurface (due to possible induced circulating currents), while the parallel impedance corresponds to the electric polarization. The degree of asymmetry of the T-circuit, proportional to the difference $Z_1-Z_2$, characterizes bianisotropic omega response~\cite{yazdi2015bianisotropic} (related to $\Omega$ parameter in (\ref{bian11}) in the bulk material case). 
	Some possible conceptual realizations of bianisotropic omega-type metasurfaces are shown in \cite[\textsection~4]{suppl}.

	As a proof of concept, here we consider the simplest circuit configuration which provides nonzero bianisotropic coupling. 
	We choose the right series circuit element as an inductor with time-invariant inductance~$L_0$, while the parallel element as a capacitor with temporally modulated capacitance $C(t)=C_0[1-A\sin(\omega_mt+\phi)]$. The left series element is short-circuited [see Fig.~\ref{fig_circuit}(a)]. 
	Based on the time-domain analysis~\cite[\textsection~5]{suppl}, the incident and transmitted voltages for forward and backward illuminations   satisfy the following relations  
	\textcolor{black}{	\begin{equation}
	v_i^f(t)=  \hat{P}(t)  v_t^f(t), \quad 
	v_i^b(t)=\left[\hat{P}(t)   +   \frac{L'}{2}\frac{d}{dt}  \frac{dC'(t)}{dt} \right] v_t^b(t),
	\label{eq_diffur_vol_forwardLc}
	\end{equation}}
	where operator $\hat{P}(t)$ is given by
	\begin{equation}
	\hat{P}(t)=1+ \frac{1}{2}\frac{ d}{ dt} \left(C'(t) +L'+ L' C'(t) \frac{ d}{ dt} \right).
	\label{eq_diffur_vol_backwardLc}
	\end{equation}
	Here, $v_i^{f,b}$ are  the incident voltage signals (equivalent to incident electric fields) for the forward and backward illuminations, $L'=L_0/\eta_0$ and $C'(t)=\eta_0 C(t)$ are the inductance and capacitance normalized by the  free-space wave impedance~$\eta_0$  with the dimensions of time. As is seen from (\ref{eq_diffur_vol_forwardLc}), the differential operators acting on transmitted voltages for the opposite illuminations $v_{ t}^{ f}(t)$ and  $v_{ t}^{ b}(t)$ differ by the term which includes time derivative of the capacitance function. 
	Therefore, if $C'(t)$ is constant, both equations in (\ref{eq_diffur_vol_forwardLc}) become identical, resulting in expected reciprocal propagation in the time-invariant metasurface. However, as will be shown below, a metasurface with nonzero ${d}C'(t)/{ d}t$ in general can exhibit  nonreciprocal transmission. 

		It is easy to test under what conditions  the metasurface described by~(\ref{eq_diffur_vol_forwardLc}) exhibits nonreciprocal propagation at frequency~$\omega_0$. To do that, we choose modulation at $\omega_{m}=2\omega_0$ \cite[\textsection~5]{suppl}. 
		\textcolor{black}{Such modulation frequency has been also applied, as examples, for wave amplification \cite{Cullen1958} and one-way beam splitting \cite{taravati2019dynamic} but using space-time modulation schemes.} 
		Here, we assume the transmission signal for both incident directions is $v_{   t}^{  f,b}=\cos(\omega_0t+\psi)$. In this way, the corresponding incident signals can be easily found by substituting $v_{ t}^{ f,b}$ into Eq.~(\ref{eq_diffur_vol_forwardLc}) \cite[\textsection~5]{suppl}. After knowing the incident fields, the transmission coefficients for forward and backward incidences at the fundamental frequency can be calculated as 
\begin{equation}
    T^{ f}(\omega_0)=4\left[Q-C_0A\omega_0(\eta_0-j\omega_0L_0)e^{j(\phi-2\psi)}\right]^{-1},\label{eq: forward}
\end{equation}	
\begin{equation}
    T^{ b}(\omega_0)=4\left[Q-C_0A\omega_0(\eta_0+j\omega_0L_0)e^{j(\phi-2\psi)}\right]^{-1},\label{eq: backward}
\end{equation}	
where $Q=4-2C_0L_0\omega_0^2+2j\omega_0(L^\prime+C_0\eta_0)$. It is obvious that $T^{ f}$ and $T^{ b}$ are not equal only if $L_0\neq 0$, which means that this structure is nonreciprocal only when bianisotropic coupling is present. 	
	\begin{figure*}
		\centering
		\includegraphics[width=0.95\linewidth]{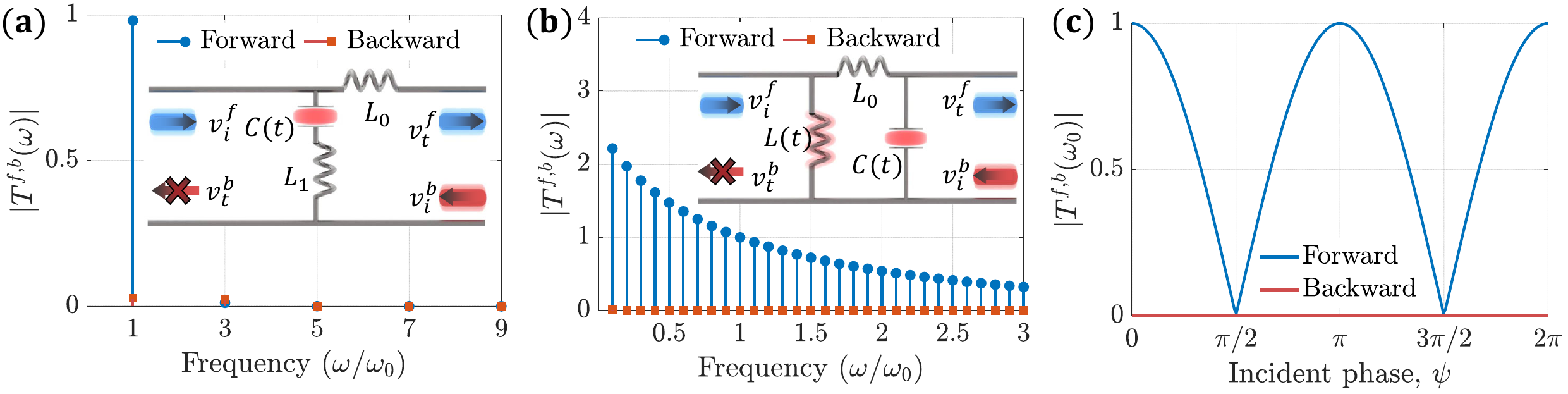}
		\caption{(a) Spectral response of the modified circuit for small modulation amplitude. Here, $C(t)=6.33[1-0.1\sin(2\omega_0 t)]$~fF, $L_0=6.2$~nH, and $L_1=38.2$~nH. (b) Spectral response of the modified circuit for reduced modulation frequency. Here, $L_0=216.1$~nH, $L(t)=149.7[1-\sin(0.1\omega_0 t)]$~nH, and $C(t)=5.36[1-\sin(0.1\omega_0 t)]$~pF.   Isolation levels achieved at $\omega_0$ is $68.1$~dB. (c) Forward and backward transmission amplitudes of the fundamental harmonic as functions of the incident phase.
		} 
		\label{fig_Csub}
	\end{figure*}	
	
	Interestingly, although the metasurface described by the circuit in Fig.~\ref{fig_circuit}(a)  is nonreciprocal, it obeys the generalized time-reversal symmetry~\cite{williamson2020integrated}. 
		Under substitution   $v(t)\rightarrow v(-t)$ and $\eta_0\rightarrow -\eta_0$ (the latter substitution is due to the reversal of the current direction in the circuit, which is defined as $i(t)=v(t)/\eta_0 $), relations (\ref{eq_diffur_vol_forwardLc}) and (\ref{eq_diffur_vol_backwardLc}) do not change their forms, providing that $C(t+\Delta t)=C(-t+\Delta t)$ for some specific gauge time translation~$\Delta t$. Therefore, signal propagation in the circuit shown in Fig.~\ref{fig_circuit}(a) obeys the generalized time-reversal symmetry \cite[\textsection~6]{suppl}. 
		Such nonreciprocal but time-reversal symmetric response was recently reported for static but non-Hermitian systems~\cite{buddhiraju_nonreciprocal_2020}. 
		Our modulated system is also non-Hermitian, i.e. energy is not conserved in the system \cite[\textsection~10]{suppl}, and nonreciprocity manifests itself in terms of  unidirectional amplification/attenuation.

The temporal modulation induces frequency mixing,  and the reflected and transmitted signals contain infinite numbers of harmonics  $\omega_n=\omega_0+n\omega_{ m}$, where $n$ is an integer and refers to the harmonic order. 
In order to choose parameters $L_0$ and $C(t)$ of the circuit providing the highest nonreciprocity at the fundamental frequency, we optimize the circuit values based on the time-Floquet analysis~\cite[\textsection~7.1]{suppl} for given incident voltages $v^{  f,b }_{  i}(t)=\cos(\omega_0 t)$. 
In the numerical optimization using MATLAB, we define the cost function,
$F=||T^{ f}(\omega_0)|-K|+|T^{ b}(\omega_0)|$,  and search for such  set of circuit parameters $\{L, C_0, A, \phi \}$ which ensures $F\rightarrow0$ \cite[\textsection~9]{suppl}. 
Parameter~$K$    defines desired transmission for the forward illumination, while for the backward illumination transmission should be always suppressed. We performed optimization of the circuit parameters for three different regimes: Forward-transmitted wave is attenuated by half (Regime I: $K=0.5$), unchanged (Regime II: $K=1$) and amplified (Regime III: $K=2$). The optimization results are shown  and confirmed with the simulated results obtained from MathWorks Simulink in Fig.~\ref{fig_circuit}(b).
The results demonstrate that the metasurface can perform one-way transmission by only modulating a single capacitor in the equivalent circuit, and the transmittance can be arbitrarily engineered with energy damping or amplification via modifying function $C(t)$. These features are very different from properties of the previously reported nonreciprocal devices~\cite{Cullen1958,dong_inducing_2008,yu_complete_2009,song2019direction,wang2020theory,qin_nonreciprocal_2014,hadad_breaking_2016,taravati2020space,taravati2017nonreciprocal,chamanara2017optical,shaltout2019spatiotemporal,shaltout2015time,cardin2020surface,zhang2019breaking,li2020time,taravati2019generalized,ramaccia2018nonreciprocity}. 
All nonzero high-order frequency harmonics can be filtered out using a conventional frequency band-pass filter. 
The power and  efficiency analysis of the system (also for the systems in Fig.~2) are presented in \cite[\textsection~10]{suppl}. From Eqs.~(\ref{eq: forward}) and (\ref{eq: backward}), it is obvious that the  nonreciprocity level can be arbitrarily tuned by adjusting the value of static inductance (more details in \cite[\textsection~11]{suppl}).  
	
Next, we implement the designed time-modulated equivalent circuit [Fig.~1(a)] using a realistic metasurface structure performing nonreciprocal transmission or amplification for plane waves. 
The parallel capacitor in   in the circuit [Fig.~1(a)] can be implemented by a array of metallic patches, as shown in the inset of Fig.~\ref{fig_circuit}(c). Under plane wave incidence, the gaps between adjacent patches exhibit capacitive property. In each gap, we embed a varactor to tune the effective capacitance of the metasurface layer. By applying a time-harmonic voltage signal on the varactors, the effective capacitance of surface will change according to the function $C(t)$.
The static inductance in the equivalent circuit  can be implemented by a dielectric substrate. 
The required bianisotropic response of this metasurface is provided by its asymmetric geometry.  
 	Applying optimization based on the time-Floquet analysis~\cite[\textsection~7.2]{suppl}, we find the optimal metasurface parameters for the three mentioned regimes (listed in the caption of Fig.~\ref{fig_circuit}(c)).  The  transmission data through the metasurface is  shown in Fig.~\ref{fig_circuit}(c). The results are similar to those in Fig.~\ref{fig_circuit}(b): 
	the metasurface blocks transmission in the backward direction but allows transmission/amplification in the forward direction at $\omega_0$. 
	
	The dynamic range of capacitance variations in the analyzed simple circuit example is relatively high, which can hinder practical implementations. Nevertheless, it can be significantly reduced by adding additional constant circuit elements to the considered circuit. In Fig.~2(a), we connect a static inductance to the time-varying capacitance in  series and optimize the modulation function to realize an isolator ($K=1$).
	The nonreciprocal effect is still evident even with the modulation amplitude as low as $A=0.1$. 
   Another issue is the high modulation speed ($\omega_{\rm m}=2\omega_0$), which can be easily realized in microwave frequencies but challenging in optics.
    However, it is important to note that in general, there is no fundamental restrictions for the choice of $\omega_{\rm m}$. Low speed modulation, such as $\omega_{\rm m}=0.1\omega_0$ and lower, can be achieved if  the equivalent circuit comprises more than one  modulated element (even having the same modulation law). 
    Figure~2(b) shows that by adding a time-varying  inductance $L(t)$ which is in-phase modulated with $C(t)$ (forming a $\Pi$-circuit), strong isolation ($K=1$) can be achieved with the modulation frequency  $\omega_{\rm m}=0.1\omega_0$. 
    Importantly, as we change the phase of the incident wave, the backward transmission is always zero while the forward transmission changes along with the incident phase. This means that, if the pumping signal is synchronized with the forward incident signal (the synchronization mechanism is conceptually shown in~\cite[\textsection~12]{suppl}), the device can perform as a  phase-insensitive isolator which can work even when   illuminated simultaneously from both sides.  
    The need for additional time-varying circuit element does not mean that one should modulate more than one components in the actual  metasurface.
    In general, the modulation of bianisotropic metasurfaces  results in time dependence of all the circuit components in their equivalent circuits \cite[\textsection~4]{suppl}.

	To summarize, we have introduced a concept of bianisotropic time-modulated systems capable of nonreciprocal wave propagation. In contrast  to other approaches for nonreciprocal systems based on temporal modulations, our route provides high isolation (or amplification) at  the  fundamental  frequency using   only uniform temporal    modulation of material properties. 
	Our  findings provide  an attractive alternative for  designing magnetless  nonreciprocal microwave devices and, under proper scheme of bianisotropic response in the metasurface, can be further extended  to higher frequencies as well as applied to waves processes of different nature.

	\begin{acknowledgments}
		This work was supported in part by the the Academy of Finland (project 309421), European Union’s Horizon 2020 Future Emerging Technologies call (FETOPEN - RIA) under Grant No. 736876 (project VISORSURF),
		the Finnish Foundation for Technology Promotion, and the U.S. Air Force Office of  Scientific Research MURI project (Grant No. FA9550-18-1-0379).
		The authors thank Dr. Momchil Minkov, Dr. Ian   Williamson,  Ms. Jiahui Wang, Prof. Andrea Al\`{u}, and Dr. Robert Duggan for useful comments and discussions about the manuscript. 
	\end{acknowledgments}


%

\end{document}